\documentclass[graphicx,11pt]{aastex}

\lefthead{}
\righthead{}

\begin{document}

\title{LATE ACCRETION AND THE LATE VENEER} 

\author{\textbf{A. Morbidelli$^{(1)}$,B. Wood $^{(2)}$}\\  
(1) Laboratoire Lagrange, UMR7293, Universit\'e de Nice Sophia-Antipolis,
  CNRS, Observatoire de la C\^ote d'Azur. Boulevard de l'Observatoire,
  06304 Nice Cedex 4, France. (Email: morby@oca.eu / Fax:
  +33-4-92003118) \\
(2) Dept. of Earth Sciences, University of Oxford, UK \\ 
} 

\begin{abstract}

The concept of Late Veneer has been introduced by the geochemical community to explain the abundance of highly siderophile elements in the Earth’s mantle and their chondritic proportions relative to each other. However, in the complex scenario of Earth accretion, involving both planetesimal bombardment and giant impacts from chondritic and differentiated projectiles, it is not obvious what the “Late Veneer” actually corresponds to. In fact, the process of differentiation of the Earth was probably intermittent and there was presumably no well-defined transition between an earlier phase where all metal sunk into the core and a later phase in which the core was a closed entity separated from the mantle. In addition, the modellers of Earth accretion have introduced the concept of “Late Accretion”, which refers to the material accreted by our planet after the Moon-forming event. Characterising Late Veneer, Late Accretion and the relationship between the two is the major goal of this chapter. 

\end{abstract}

\section{Introduction}

Late Veneer, Late Accretion and Late Heavy Bombardment are concepts introduced by different scientific communities to address the very last stage of Earth formation when our planet acquired the final bit of its mass, after the end of the core-mantle differentiation process and/or the giant impact that gave origin to the Moon. Often, however, the same names are used to indicate quite different processes. Moreover, it is quite common to identify the Late Veneer/Accretion/Bombardment with the origin of volatiles (including water) on our planet, which is not necessarily correct. 

Therefore, the goal of this chapter is to define these concepts, describe the processes that they refer to and their mutual relationships. We start in Section 2 with the Late Veneer, as defined by the geochemical community. In Section 3 we briefly describe the mode of formation of the Earth, which leads in a natural way to define the concept of Late Accretion. Section 4 will discuss the relationships between Late Veneer and Late Accretion. Finally in Section 5 we will come to the issue of the delivery of volatiles to the Earth and its chronology.

\section{The Late Veneer as defined in geochemistry}

The concept of a “Late Veneer” was developed from the observed abundance patterns of siderophile (iron-loving) elements in the silicate Earth. Figure 1 shows the relative abundances of a wide range of elements (McDonough and Sun 1995) in the primitive upper mantle (otherwise known as bulk silicate Earth, BSE) as a function of their condensation temperature from a gas of solar composition at a total pressure of $10^{-4}$ bar (Lodders 2003). Refractory lithophile elements, which condense at high temperatures, (e.g Ca, Ti, Sc, Zr,REE)  are in approximately the same ratios one to another as in the CI chondrite class of primitive meteorites. This gives us a reference level for all other elements. Refractory siderophile elements are depleted in BSE due to their partitioning into the core. To a first approximation one can make a mass balance between the BSE and the “missing” contents (assuming chondritic ratios in bulk Earth) of these siderophile elements and obtain the mass of the core and its elemental concentrations (dominated by Fe). Additionally, as can be seen in Figure 1, those elements with low condensation temperatures, which were volatile in the solar nebula, are depleted in BSE relative to the chondritic reference. This is due, as discussed in section 5, to the Earth being dominated by high temperature materials which condensed and accreted in the inner solar system. Note that the same relationship between siderophile and lithophile elements observed in refractory elements applies to the more volatile elements: volatile siderophile elements are more depleted in bulk silicate Earth than volatile lithophile elements.

\begin{figure}[t!]
\centerline{\includegraphics[height=8.cm]{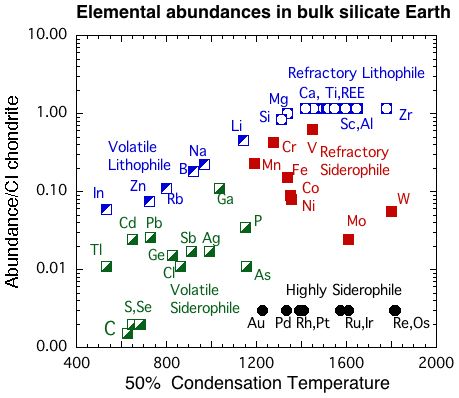}}
\caption{\small Abundance relative to CI chondrites and Mg of elements in the bulk silicate Earth, as a function of condensation temperature. The latter is defined as the temperature at which 50\% of the element is in solid form, at the pressure conditions on the Minimum Mass Solar Nebula (Lodders, 2003).  }
\end{figure}

Measurements of the concentrations of the noble metals (Re, Os, Ir, Ru, Pt, Rh, Pd, Au) in mantle peridotites and igneous rocks generated by partial melting of the mantle (Chou 1978; Chou, et al. 1983) demonstrated that these “highly siderophile” elements (HSEs) are, as expected, even more depleted in BSE than the other important siderophile elements shown in Figure 1. Experimental measurements show, however, that despite their strong depletions, these elements are much more abundant in silicate Earth than would be expected from simple equilibrium between mantle and core (Holzheid, et al. 2000; Kimura, et al. 1974; Mann, et al. 2012; O'Neill 1991). In other words, they are orders of magnitude more siderophile than indicated by their relative abundances in BSE. Partition coefficients of elements depend on temperature and pressure (Righter, et al. 2008) so that, in principle, it is possible that the values measured in laboratory experiments do not correspond to the conditions existing on Earth during core formation. However, HSEs are in approximately chondritic ratio one to another in the primitive upper mantle (Chou 1978) and it is highly unlikely that conditions of temperature and pressure exist to make the partition coefficients of the 8 HSEs equal to each other. In fact, no combination of pressure, temperature and oxygen fugacity appears to be capable of reproducing the mantle concentrations of the HSEs by metal-silicate equilibrium (Mann et al. 2012).

Thus, the most plausible explanation for the HSEs’ overabundance in BSE and their chondritic relative abundances is  the addition of material of approximate chondritic composition to the upper mantle after core segregation had ceased (Chou 1978; Kimura, Lewis and Anders 1974). These are the origins of the concept of the “Late Veneer”, a small amount ($<1$\% Earth mass) of chondrite-like material added to a convecting silicate Earth and never re-equilibrated with the metal in the core (Anders 1968; Turekian and Clark 1969). From the abundance of HSEs in BSE,  the mass of the mantle and the range of HSE contents in the various types of chondritic meteorites, the modern estimate for the Late Veneer mass is  $4.86 \pm 1.63\times 10^{-3} M_E$, where $M_E$ is the Earth's mass (Walker, 2009; Jacobson et al., 2014).   

\begin{figure}[t!]
\centerline{\includegraphics[height=8.cm]{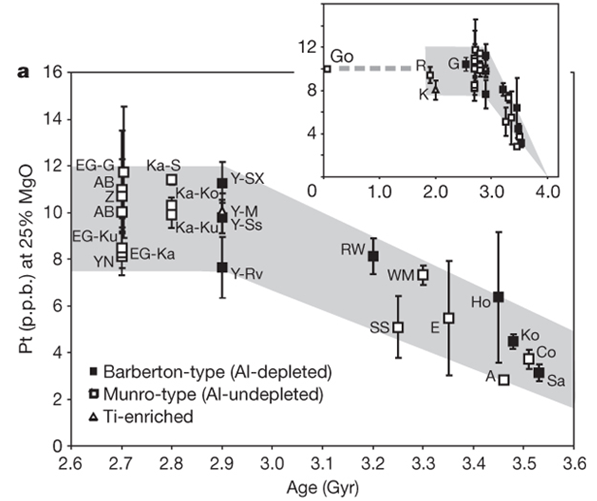}}
\caption{\small The abundance of HSEs as a function of age for the Earth’s oldest komatiites. From Maier et al., 2009. }
\end{figure}

It is important to emphasize that, as far as the highly siderophile elements are concerned, the late chondritic veneer has been completely homogenized into the mantle by convection. This process of mixing might have taken a long time to be completed. Indeed, the HSE content in komatiites decreases with their age, for komatiites older than 3 Gy (see Fig. 2; Maier et al., 2009; Wilbold et al., 2011), with one exception pointed out in Touboul et al. (2012) . It seems unlikely that this trend could be due to a slow delivery of HSEs to the Earth, because the terrestrial bombardment should have decayed much faster than is implied by this HSE vs. age trend (see Section 3). So, the most likely interpretation is that it took about 1.5 billion years to completely mix the mantle and bring the HSEs from near the surface to the deeper parts of the mantle from where the old komatiites are derived. 

\section{Late Accretion mass as defined in accretion models}

The detailed description of the process of terrestrial planet growth is the object of the chapter by Jacobson and Walsh in this book. Here it is enough to emphasize that terrestrial planets are believed to grow from a disk of planetesimals (small bodies resembling the asteroids) and planetary embryos (bodies with a mass intermediate between the mass of the Moon and the mass of Mars, which formed very rapidly, before the disappearance of gas from the system). In view of its short accretion timescale, Mars itself could be a stranded planetary embryo (Kleine et al., 2004b; Halliday et al., 2005; Dauphas and Pourmand, 2011; Jacobson and Morbidelli, 2014). 

Therefore, the growth of a planet like the Earth is characterized by a sequence of giant impacts from planetary embryos and smaller impacts from planetesimals. In this contest, we define “Late Accretion” as the tail of the accretion history of a planet due to the planetesimal bombardment which happens after the last giant impact has occurred. For the Earth, the last giant impact is presumably the one which generated the Moon (we will provide evidence for this in Sect. 4.3), so that Late Accretion occurs after lunar formation.  

Because the dynamical lifetime of planetesimals is finite (planetesimals are removed by collisions with the planets, ejection to hyperbolic orbit, collisions with the Sun or mutual collisions which grind them into small debris), the number of planetesimals in the system decays roughly exponentially with time. Hence the bombardment rate of the planets is expected to decay in the same fashion. From numerical experiments, the half-life of the process is from 10 to 50 My (Bottke et al., 2007; Morbidelli et al., 2012). Unfortunately, geological activity on the Earth has erased any clear evidence of impacts occurring over the first billion years or so. Instead, a lot of information is preserved in the lunar ancient crater record. 

From lunar craters we learn that the size distribution of the projectiles striking the Moon was similar to that of today's main belt asteroids (Strom et al., 2005). The so-called impact basins (“craters” with diameters larger than 300km) trace the impacts of the largest projectiles (up to $\sim$200km in diameter for the projectile of the SPA basin). The last of the basin-forming events occurred relatively late compared to the time of formation of the first solids in the Solar System (the Calcium-Aluminium Inclusions, which formed 4.567 Gy ago; Bouvier and Wadhwa, 2010). The Imbrium basin which, from stratigraphy analysis, appears to be the second from last basin in the chronological sequence (Wilhelms, 1987) formed 3.85 Gy ago according to the dating of lunar samples returned by the Apollo program (Stoffler and Ryder, 2001). The Orientale basin (the last basin on the Moon) presumably did not occur long after Imbrium, given that crater counts on the ejecta blankets of both basins are similar. 

Forming basins so late as a tail of an exponentially decaying process would require that the initial population of large planetesimals was very numerous (Bottke et al., 2007). However, this is inconsistent with the small amount of mass accreted by the Moon since its formation, which is of the order of $5\times 10^{-6}M_E$ as deduced from HSEs abundance in the lunar mantle (Day et al., 2010) and in the crust (Ryder, 2002) and the total number of lunar basins (Neumann et al., 2013). This argues strongly in favor of a surge in the frequency of impacts sometime prior to the formation of Imbrium (Bottke et al., 2007). Several other pieces of evidence point to such a surge, often called lunar Cataclysm or Late Heavy Bombardment (the latter term, however, is sometimes associated with the entire Late Accretion process, so we will not use it in this review to avoid confusion). For instance, impact ages on lunar samples (Tera et al., 1974) and lunar meteorites (Cohen et al., 2000) indicate that the impact rates were lower before $\sim$4.1 Gy than in the subsequent 3.9-3.7Gy period. 

A surge in the bombardment rate, associated with an increase of the velocity distribution of the projectiles, is consistent with a sudden change in the orbital configuration of the giant planets, which would have dislodged part of the asteroid population from the main belt, refurbishing the population of planet-crossing objects (Bottke et al., 2012). Such a change is needed, sometime during the history of the solar system, to reconcile the current orbits of the giant planets with those that these planets should have had when they emerged from the phase of the gas-dominated proto-planetary disk (Morbidelli et al., 2007; Morbidelli, 2013). 

Observational constraints (Marchi et al., 2012) and theoretical modeling (Bottke et al., 2012) suggest that about 10 basins formed on the Moon during the cataclysm, from Nectaris to Orientale and that the cataclysm started about 4.1 Gy ago (see Morbidelli et al., 2012 for a review). The total number of basins on the Moon, detected by the GRAIL mission through gravitational anomalies (Neumann et al., 2013) is $\sim$40. Thus, the cataclysm accounts for about 1/4 to 1/3 of the total mass ($\sim 5\times 10^{-6}M_E$) that the Moon accreted since its formation. From all these constraints, a timeline of the lunar bombardment can be derived (Morbidelli et al., 2012).

Knowing the impact history of the Moon, one can then scale it to the Earth, using the ratio of gravitational cross sections. Monte Carlo simulations also allow assessments of the effects of small number statistics related to the largest impactors (Marchi et al., 2014). The results change vastly depending on whether one limits the size frequency distribution (SFD) of the impactors to the size of Ceres (the largest asteroid in the main belt today, $\sim$900km in diameter) or extrapolates it to larger sizes. 

If the projectile size distribution is limited to Ceres-size, given the constraint on the mass delivered to the Moon, the total amount of mass that the Earth acquires during Late Accretion is less than $10^{-3} M_E$, significantly smaller than the chondritic mass estimated for the Late Veneer (see Section 2). But if planetesimals larger than Ceres are permitted, the same constraint on the mass delivered to the Moon allows a significant fraction of the simulations ($\sim$12\%) to bring a mass to the Earth consistent with the mass of the Late Veneer (see Fig. 3 and Bottke et al. 2010, Raymond et al., 2013). About 8\% of the simulations bring a somewhat larger mass. However, only 0.5\% of the simulations deliver more than 0.01$M_E$ to the Earth.  

\begin{figure}[t!]
\centerline{\includegraphics[height=11.cm]{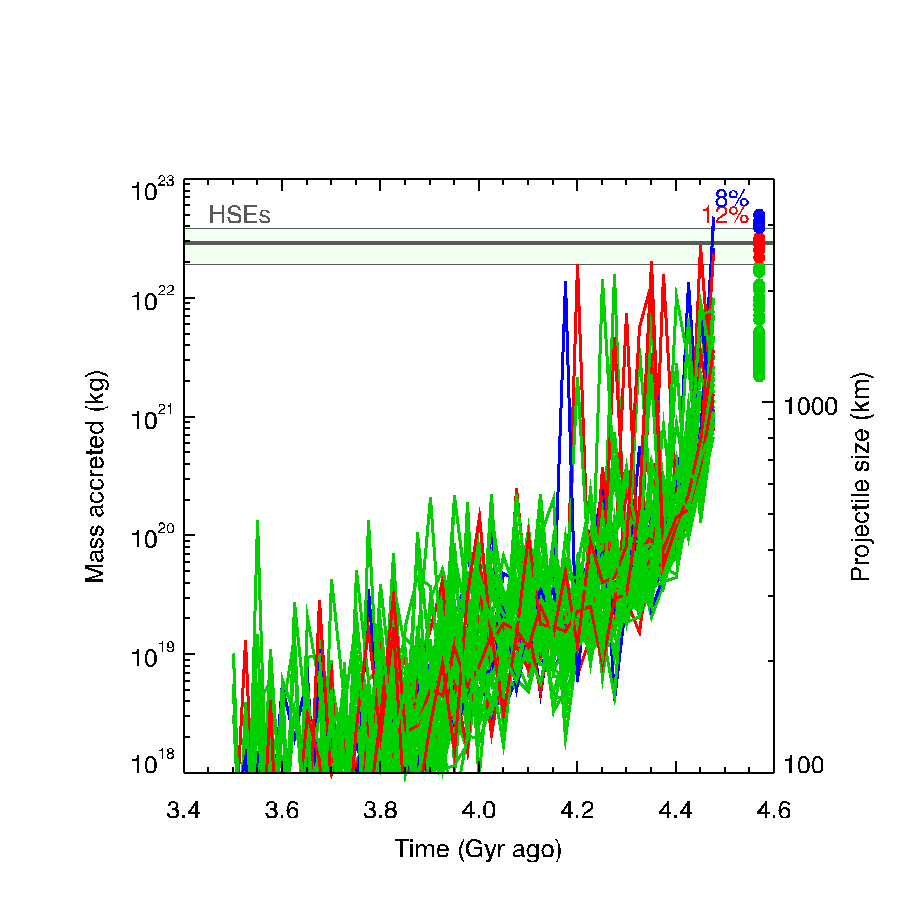}}
\caption{\small The bombardment of the Earth obtained in 50 Monte Carlo simulations calibrated on the bombardment history of the Moon. Prior to 4.1 Gy ago, the  asteroid-like SFD of the impactors is extrapolated up to $D=4000$km projectiles. Since 4.1 Gy ago, the projectile SFD is limited at Ceres’ size. Each curve shows the mass accreted in each 25My time-bin. The cumulative accreted mass is indicated by the dots on the right hand side of the panel. The horizontal light green band marks the Late Veneer mass and its uncertainty. Res curves (12\% of the runs) cumulatively carry a mass consistent with the Late Veneer mass, green curves (80\%) carry a smaller mass and blue curves (8\%) a larger mass. Adapted from Marchi et al., 2014.  }
\end{figure}

\section{Relationship between Late Veneer and Late Accretion }

As we have seen above, the Late Veneer mass and the Late Accretion mass are defined from two different concepts. The amount of Late Veneer mass is well constrained, but it may be questionable if all of it was delivered to the Earth after the Moon-forming event. The estimate of the Late Accretion mass is quite uncertain, as it depends on the assumed size distribution of projectiles when scaling the lunar impact crater record to the Earth. It would be very useful if we could equate with confidence the Late Accretion mass to the Late Veneer mass. To assess whether this equation is reasonable, we address two key questions below

\subsection{Can the Late Accretion mass be significantly smaller than the Late Veneer mass?}

To answer this question we need to consider whether any of the HSE content of bulk silicate Earth could have accreted prior to the Moon forming event. The Moon is believed to have been formed during the last giant impact on Earth. Because of its large size (at least more massive than 1/4 of Mars), the projectile is expected to have been differentiated. Numerical simulation of the Moon forming process (Canup and Asphaug, 2001; Reufer et al., 2012; Cuk and Stewart, 2012; Canup, 2012) show that the core of the projectile merges with the core of the Earth and that the temperature of the Earth is raised above the global magma ocean threshold. However, it is unclear which fraction of the terrestrial mantle has a chance to equilibrate with the projectile’s core in this process. Also, it is unclear whether there is still metallic iron in the Earth’s mantle at this stage that can take advantage of the new magma ocean phase to percolate into the core.  Obviously, if a substantial fraction of pre-existing mantle HSEs had not been depleted during the episode of core growth associated with the Moon-forming event, the Late Veneer mass deduced from the current HSE abundance would overestimate the mass accreted after the Moon’s formation, i.e. the Late Accretion mass. 

If we consider the abundances of the refractory siderophile elements Ni and Co in BSE (Fig 1), then it has been known for nearly 20 years that these can only be explained, under conditions of metal-silicate equilibrium, by core segregation at very high pressures (Li and Agee 1996; Thibault and Walter 1995). When the weakly siderophile elements V, Cr and Nb are also considered, the most likely explanation of the core formation process is that it began under strongly reducing conditions at low pressure and temperature and that the Earth became more oxidized as it grew and the pressures and temperatures of core segregation increased (Wade and Wood 2005; Wood, et al. 2008). These observations led Mann et al (2012) to determine the pressure, temperature and oxygen fugacity effects on the metal-silicate partitioning of the HSE’s (Re, Ir, Ru, Pt, Rh, Pd). Their data demonstrate that these elements all become less siderophile with increasing temperature and to some extent with increasing pressure and that by a long extrapolation one might infer that the mantle concentrations of Pt, Rh and Pd would approach those observed at equilibrium conditions of about 60 GPa and 3560K. Nevertheless, the metal-silicate partition coefficients of Ru, Re and Ir would remain 1-2 orders of magnitude too high to explain the mantle concentrations of these elements even under these extreme conditions. 

Since equilibrium between core and mantle would not generate the observed abundance pattern of HSE’s in BSE, the only way in which the current concentrations could contain some “pre-Moon-formation” component is if it were a result of partial disequilibrium during the Moon-forming event. For example, a small amount of metal left behind by inefficient core segregation would contain the HSE’s in chondritic relative proportions because these elements are so strongly partitioned into the metal. This would be the case, for instance, if the Earth’s hemisphere opposite to the impact point of the Moon-forming projectile remained largely unaffected by the collision, as suggested by the recent simulations in Cuk and Stewart (2012) (where the mantle in the opposite hemisphere becomes the lower mantle of the final Earth -- although its temperature is very high so that it could still undergo metal-silicate segregation). 

Alternatively, internal production of Fe$^{3+}$ in the mantle by, for example, disproportionation of Fe$^{2+}$ to Fe$^{3+}$ plus Fe$^0$ in the perovskite field (Frost, et al. 2004; Kyser, et al. 1986) could lead to the oxygen fugacity of the upper mantle exceeding the conditions of iron metal stability. Any additional metal added (including the HSE’s) would then be oxidized and mixed back into the silicate part of Earth rather than being segregated to the core. In this way the giant impact associated with the Moon-forming event would have delivered HSEs to the Earth’s mantle, rather than depleting them. 

Tungsten isotopes provide a potential means of independently estimating both the mass of the Late Veneer and of Late Accretion. $^{182}$W was produced in the early Earth by decay of $^{182}$Hf with a half-life of 8.9 My. This extinct radioactive system, in which W is siderophile and Hf is lithophile is extensively used to estimate the accretionary timescales of planetary bodies (Yin, et al. 2002). 

A Late Veneer which provided $\sim 0.5$\% of an Earth mass with HSEs in chondritic form should have provided the same amount of W with chondritic $^{182}$W/$^{184}$W i.e. with $\epsilon_{\rm W}$ of -1.9. Given that siderophile W was strongly partitioned into the core during the principal accretionary phase and depleted by a factor $\sim$16 in the mantle relative to chondrites, this amount of chondritic W is sufficient to have significantly altered the $\epsilon_{\rm W}$ of BSE, decreasing it by about 0.25 in $\epsilon_{\rm W}$  units. Measurements of the W isotopic compositions of 3.8 Gyr age rocks from Isua, Greenland (Willbold, et al. 2011) show that these have slightly higher $^{182}$W/$^{184}$W ($\epsilon_{\rm W} = 0.13 \pm 0.04$) than modern day silicate Earth ($-0.01 \pm 0.02$). This result is consistent with the Isua rocks reflecting the composition of BSE prior to the Late Veneer (this interpretation should be validated by checking that Isua rocks are HSE poor), while the modern Earth exhibits homogenization of the Late Veneer with the pre-existing mantle (see Fig. 2). Willbold et al (2011) show that this shift in the W isotopic composition of BSE, if attributed to mixing-in of the Late Veneer, would lead to a mass of the Late Veneer of between 0.002 and 0.009 of the mass of bulk silicate Earth, depending on the nature of the chondritic material added in the Late Veneer. Thus, W-isotopes yield an estimate for the mass of the Late Veneer which is consistent with all HSE’s in silicate Earth arriving with the Late Veneer. This argument however does not require that this mass was delivered after the Moon forming event.

For this purpose it is interesting to look at the differences in W isotope composition between the Earth and the Moon. The Earth and the Moon have essentially identical isotope compositions for many elements: O (Wiechert et al., 2001), Ti (Zhang et al., 2012), Si (Georg et al, 2007), Cr (Lugmair and Shykolyukov, 1998). Whatever the origins of these identical isotopic compositions, it is reasonable to assume that, at formation, the Moon and the Earth had identical W-isotope composition as well. However, if the HSE abundances are diagnostic of Late Accretion, the Earth should have subsequently accreted much more chondritic material than the Moon. Thus, its W-isotope composition should now be different from that of the Moon; the Moon should now have $\epsilon_{\rm W} \sim 0.2$. Until recently, the W-isotope compostion difference between the Earth and its satellite was not resolved: $\epsilon_{\rm W} ({\rm Moon}) = 0.1\pm 0.1$ (Touboul et al., 2007). Recently, however, Kleine et al. (2014) reported to have achieved a more precise determination: $\epsilon_{\rm W} ({\rm Moon}) = 0.29 \pm 0.08$ (the value reported in the reference is 0.17 but it has been later refined to 0.29 --  Kleine, private communication). Originally Touboul et al. interpreted any difference in W-isotopic composition between the Moon and the Earth as an indication that the Moon formed before all $^{182}$Hf was extinct.  However, it now turns out that the Hf/W ratios in the Earth and Moon mantles are the same (K\"onig et al., 2011) so the W-isotope composition of the two bodies should not change relative to each other over time, whatever the age of the Moon. Thus, the difference reported by Kleine et al. is more easily explained if a chondritic mass consistent with the Late Veneer mass was delivered to the Earth after the Moon forming event. 

If W suggests that the Late Accretion mass cannot be significantly smaller than the Late Veneer mass, two arguments suggest that nevertheless a fraction of the HSEs should pre-date the closure of Earth's core and/or the Moon forming event. 

A first argument comes from sulfur isotopes. As can be seen in Figure 1, S is depleted in BSE due to its volatility in the solar nebula and also because it is highly siderophile, even though it is not conventionally named as one of the HSE’s. To a first approximation, the S content of BSE (250 ppm, McDonough and Sun, 1995) is consistent with an addition of 0.5\% of CI chondrite to the Earth in the Late Veneer. Thus, its abundance is consistent with the Late Veneer mass. However, recent measurements of the S-isotopic composition of BSE (Labidi, et al. 2013) have shown that the dominant mantle end-member of S sampled by MORB has $\delta^{34}$S of -0.19\% rather than the value of $\sim$0.0, expected for chondritic sulfur. Measurements of the S isotopic compositions of carbonaceous, ordinary and enstatite chondrites range from -0.026 to +0.049 \% (Gao and Thiemens 1993) so the mantle end-member is compositionally distinct from conventionally recognized planetary “building-blocks”. Based on limited experimental data on S isotope partitioning between silicate and metal, Labidi et al (2013) showed that the difference between chondrites and the mantle end-member is consistent with the presence of sulfur in the mantle which had equilibrated with the metal of the core. The isotopic measurements are consistent with the amount of “Late-Veneer” sulfur in the mantle being about 40\% of the total, although this figure has very large uncertainties. The conclusion is that some fraction of highly siderophile, volatile S was present in the mantle before the Late Veneer was added and that it is probable that the same applies to other highly siderophile elements as well. 

A second argument is brought by the analysis of Kostomuksha komatiites (Touboul et al., 2012), for which combined $^{182}$W,$^{186,187}$Os, and $^{142,143}$Nd isotopic data indicate that their mantle source underwent metal--silicate fractionation well before 30 Myr of Solar System history. Surprisingly, Kostomuksha komatiites appear to come from a mantle source which had a HSE content which was 80\% of that of the present-day mantle. Thus, the fraction of the current HSE mantle budget that predates the Moon forming event would be $0.8 \times f$, where $f$ is the fraction of the total mantle mass represented by the source of the Kostomuksha komatiites. Unfortunately, $f$ is not known even at the order of magnitude level

In summary, we can exclude the Late Accretion mass being order(s) of magnitude less than the Late Veneer chondritic mass deduced from HSE abundances in the terrestrial mantle. For instance we can exclude that, after the Moon forming event, the Earth accreted only 0.0001 $M_E$, which would be the mass expected by scaling the mass accreted by the Moon ($5\times 10^{-6}M_E$) according to the gravitational cross-sections. This supports the scenario of stochastic accretion proposed by Bottke et al. (2010), i.e. that the projectile size distribution extended to sizes significantly larger than Ceres (see the right panel of Fig. 3). However, it is possible and plausible that a fraction of the HSEs, perhaps as much as $\sim$50\%, pre-date the Moon-forming event. 

\subsection{4.2	Can the Late Accretion mass be significantly larger than the Late Veneer mass?}

If the impactors delivering most of the mass during Late Accretion were very big (larger than 1000km in diameter), as advocated by Bottke et al. 2010, they were presumably differentiated objects. Thus, most of their HSEs were sequestered in their cores. The idea proposed in Bottke et al. is that the cores of these objects equilibrated with the terrestrial mantle, the metallic iron was oxidized and therefore it did not join the terrestrial core, so that the projectile’s HSEs were delivered to the Earth’s mantle. However, if only a fraction $X$ of the projectiles’ cores equilibrated with the mantle, while the remaining $1-X$ fraction plunged directly into the Earth core, the HSE mantle budget would require a Late Accretion mass $M_{LA}$ equal to $M_{LV}/X$, where $M_{LV}$ is the Late Veneer mass. Thus, Albarede et al. (2013) suggested that, if $X$ is small, $M_{LA}$ can be much larger than $5\times 10^{-3} M_E$. 

An upper bound on the late-accreted mass of Earth can be derived from the fact that the silicate parts of Earth and Moon have very similar isotopic compositions. If a large mass had been added to Earth in a few large projectiles after the Moon forming event, the chemical nature of Late Accretion would have been dominated by a single parent body composition (rather than an average of many projectile properties) and then the Earth and the Moon would be more different than they are observed to be. Below, we examine the isotopic systems of O and Ti.

Wiechert et al. (2001) found that Earth and the Moon lay on the same O isotope fractionation line:
$$
 \Delta ^{17}{\rm O(Moon)}- \Delta ^{17}{\rm O(Earth)} = 0.0008 \pm 0.001{\rm \%}
$$
where $\Delta ^{17}O$ measures the distance from the terrestrial fractionation line on an oxygen three-isotope diagram. All reported uncertainties are 1$\sigma$. Recently, Herwartz (2013) announced that they resolved a difference between the two bodies: $\Delta ^{17}{\rm O(Moon)} - \Delta ^{17}{\rm O(Earth)} = 0.0012$\%. If this measurement is correct and if we assume that Earth and Moon equilibrated oxygen isotopes at the time of the Moon-forming event, knowing that the Earth received more late-accreted mass than the Moon, we conclude that the dominant nature of the projectiles during Late Accretion must have had a carbonaceous chondrite (excluding CI), enstatite or HED meteoritic composition, because these are the only meteoritic compositions that are situated below or on the terrestrial fractionation line in the oxygen-isotope diagram. Then we can constrain how much late-accreted mass could be added to Earth from carbonaceous CV, CO, CK, CM, CR and CH meteoritic compositions using their measured oxygen isotope compositions. We find: $0.4\pm 0.3$\%, $0.3\pm 0.2$\%, $0.3\pm 0.2$\%, $0.5\pm 0.5$\%, $0.8\pm 0.7$\% and $0.8\pm 0.7$\% Earth mass, respectively. For HEDs and enstatite chondrites, oxygen isotopes do not provide very useful constraints, because these meteorite groups are close to the terrestrial fractionation line.

Further constraints are provided by the relative Ti isotope composition of the Earth and the Moon. The Moon–Earth difference is:
$$
\epsilon ^{50}{\rm Ti(Moon)} - \epsilon ^{50}{\rm Ti(Earth)} = -0.04\pm 0.02
$$ 
(Zhang et al., 2012). HEDs and, to a lesser extent, enstatite chondrites, can be excluded as the dominant component of Late Accretion because they have negative titanium isotope compositions ($\epsilon ^{50}{\rm Ti(enstatite)} = -0.23\pm 0.09$ and $\epsilon ^{50}{\rm Ti(HED)} = -1.24\pm 0.03$) and any large addition of this composition would result in a positive $\epsilon ^{50}$Ti measurement difference between the Moon and Earth. A source of uncertainty in this analysis appears from the possibility that the Moon–Earth difference is positive, because its 2$\sigma$ uncertainty encompasses the zero value. Considering this possibility and assuming that $\epsilon ^{50}{\rm Ti(Moon)} - \epsilon ^{50}{\rm Ti(Earth)} = 0.01$ (a 2.5$\sigma$ deviation), we can still constrain that at most $0.008 M_E$ of HED composition could have been added to the Earth; however, we cannot place such a similarly strict constraint on a projectile of enstatite chondrite composition. 

\begin{figure}[t!]
\centerline{\includegraphics[height=7.cm]{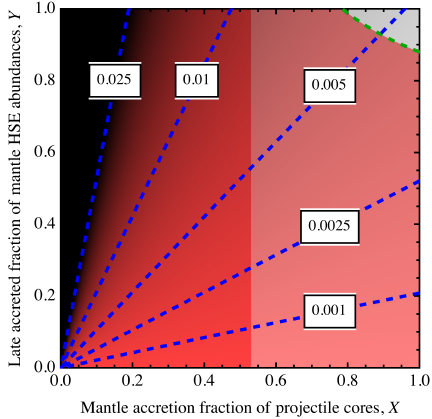}\quad \includegraphics[height=7.cm]{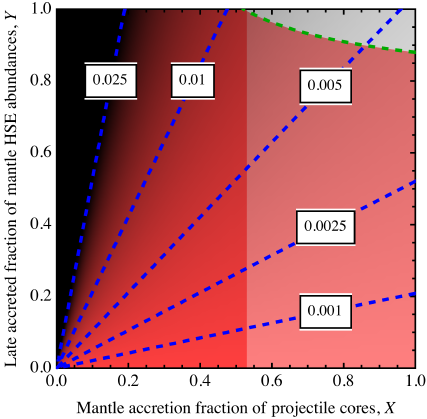}}
\caption{\small In the parameter space $X$ (fraction of the projectile core that equilibrates with the mantle) and Y (fraction of HSEs in the silicate Earth delivered as Late Accretion) the gray region represents the parameter space consistent with a value of $\epsilon$W (Moon-Earth)= $0.29\pm 0.08$ (Kleine et al. 2014). The left panel has been computed for a Eucrite-like composition of the mantle of the differentiated projectile and the right panel for an Aubrite-like composition (in terms of W-content and $\epsilon$W). The diagonal dashed blue lines show the total mass accreted by the Earth. Adapted from Jacobson et al. (2014) using the new value of the lunar $\epsilon$W and a mass concentration of W in the Earth’s mantle that is 1/16 of chondritic. }
\end{figure}

In order to constrain the mass that the Earth could have accreted from differentiated enstatite chondrite projectiles we turn again to the difference in W-isotope ratio between the Moon and the Earth. If only a fraction $X$ of the cores of the projectiles equilibrated with the mantle, a fraction $1-X$ of the projectiles’ mantles have to have been delivered as an achondritic contribution to the terrestrial mantle. Because enstatite chondrites are very reduced, the mantle of differentiated enstatite chondrite objects is expected to be very depleted in W, but with a highly radiogenic isotopic ratio (i.e. large positive $\epsilon$W). Eucrite and Aubrite meteorites could be used as a proxy of what the mantle of a differentiated enstatite-chondrite projectile could look like ($\epsilon$W equal to 22 and 11, respectively). This radiogenic contribution would increase $\epsilon$W of the BSE relative to the Moon, making the Moon negative in $\epsilon$W relative to the final terrestrial standard (which sets the zero value by definition). But we have seen before that the Moon has a positive $\epsilon$W=0.29 relative to the BSE. This effectively limits the amount of mass delivered by differentiated enstatite chondrite projectiles. 

Fig. 4 shows this constraint in graphical form. Depending on the assumed composition of the projectile mantle (Eucrite-like -- left panel or Aubrite like -- right panel), the figure shows as a gray area the set of parameters $X$, $Y$ ($Y$ being the fraction of HSEs in the BSE delivered after the Moon forming event) consistent with the measured Moon-Earth difference in $\epsilon$W and it’s uncertainty. The diagonal dashed blue lines show the total mass delivered to the Earth. 
 
As one can see, also in this case the amount of material delivered to the Earth should not exceed 0.01~$M_E$.

\subsection{Summary and implications for Moon formation.}

The discussions in Sects. 4.1 and 4.2 suggest that the Late Veneer mass and the Late Accretion mass are the same within a factor of 2. In particular, the Earth should not have accreted more than 1\% of its mass after the Moon-forming event. This result has two implications.

\begin{figure}[t!]
\centerline{\includegraphics[height=7.cm]{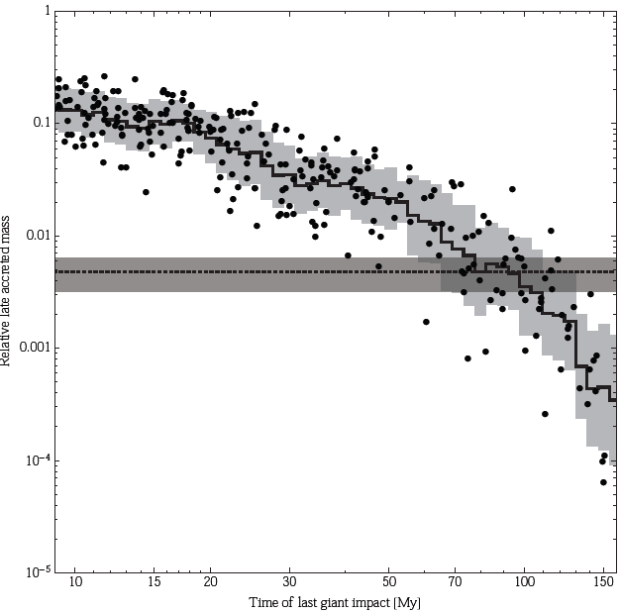}}
\caption{\small  Each dot represents a planet produced in a terrestrial planet formation simulation, indicating the time of the last giant impact and the amount of mass (normalized by the final planet mass) accreted from planetesimals after that impact. The distribution of points shows a clear correlation whose running mean and standard deviation are illustrated by the solid staircase-like line and the surrounding gray band, respectively. The Late Veneer mass and its uncertainty are depicted as a horizontal dark gray band.  Adapted from Jacobson et al. (2014).}
\end{figure}

First, the Moon should have formed during the last giant impact that brought mass to the Earth. In fact, the short accretion timescale of Mars suggests that planetary embryos had masses of the order of 0.5-1 Mars mass. Thus, each accretional giant impact of embryos on Earth should have delivered at least 5-10\% of an Earth mass to our planet. If the Moon-forming giant impact had not been the last one, our planet would have accreted several percent of its mass during the subsequent giant impact, which is excluded by the arguments presented in Sec. 4.2.

Second, the Moon formed relatively late. Jacobson et al. (2014) showed with computer simulations that there exists a correlation between the timing of the last giant impact and the mass that the planet acquires afterwards (Fig. 5). From this correlation one deduces that the Moon should have formed from 65 to 120My after formatiom of the first solids for to Earth to accrete a Late Veneer of 
$4.86 \pm 1.63 \times 10^{-3} M_E$. If the Late Accretion mass is 0.01 $M_E$, the Moon should have formed no earlier than 50My, within 1$\sigma$ uncertainty. This is in agreement with the date of the Moon-forming collision using some radioactive chronometers (e.g. Allegre et al., 2008 using the Pb-Pb chronometer; Halliday, 2008 using the Rb–Sr chronometer), but it is well known that other radioactive chronometers are discordant (e.g. Yin and Antognini, 2008 for the Hf-W chronometer; Taylor et al., 2009 for the Lu-Hf chronometer). 

\section{Late Veneer and the origin of Earth’s volatiles}

An important issue which has arisen recently is the relationship between Late Veneer and the origin of terrestrial volatiles, including moderately volatile elements such as S, Pb and Ag (Fig 1). Cosmochemically the simplest model for the origin of volatiles is that the Earth accreted from volatile-depleted materials which changed little in bulk composition as accretion progressed (homogeneous accretion). An alternative hypothesis (Albarede, 2009) is that the Earth first formed completely devoid of volatile constituents.  Albarède suggests that the Earth’s complement of volatile elements could reasonably have been established later, by addition of volatile-rich material after the Moon-forming event.

There are several chemical and isotopic arguments in favour of a view intermediate between the homogeneous accretion model and the extreme heterogeneous accretion model of Albarede (2009). Firstly, although the silicate Earth is substantially depleted in volatile elements relative to CI chondrites (Fig. 1) it is clearly most strongly depleted in those volatile elements which are also siderophile (e.g., S, Se, Au, Ge, C). This suggests that these elements have partitioned into the core, implying that volatile elements were added to Earth during the principal phase of accretion and core segregation, and not solely as part of a Late Veneer. S-isotopic evidence indicating that some fraction of terrestrial S equilibrated with metal of the core (Labidi et al 2013) is further evidence for a pre-Late Veneer volatile accretion to the Earth. 

A second argument against the accretion of volatile elements solely in the Late Veneer is that such a model can hardly be mass-balanced. For example the concentration of $^{204}$Pb in CI chondrites (the most volatile rich meteorites) is 42 ppb (Palme and O'Neill 2003). In contrast, the silicate Earth contains 2.5 ppb of $^{204}$Pb  (Fig. 1). Supplying this as a Late Veneer of CI chondrite composition, would add 6\% to the mass of the primitive mantle and deliver Re, Ag, S, C and water at levels that are 5, 7, 11, 13 and 5 to 20 times greater, respectively, than found in bulk silicate Earth. Atmospheric losses may reduce the final abundances, but unlikely in the required proportions. Delivering all terrestrial Pb during the Late Veneer would also generate a Moon-Earth difference in W isotopic composition of $\sim$ -1.3$\epsilon$W and of 0.05\% for $\Delta ^{18}$O, significantly different from observations (Touboul, et al. 2007; Wiechert, et al. 2001). These inconsistencies are made even worse if the Late Veneer were more volatile-depleted than CI chondrites. For example, Re-Os systematics indicate that the Late Veneer may have had a composition similar to that of H chondrites (Drake and Righter, 2002). In this case, the $^{204}$Pb mass balance requires addition of $\sim$60\% of the mass of the Earth’s mantle! 

A complement to these arguments against the Late Veneer as the sole source of the volatile elements is provided by Schonbachler et al. (2010). These authors show that the silicate Earth is identical to CI chondrites in Ag isotopes. Since $^{107}$Ag was produced in the early solar system by decay of $^{107}$Pd ($t_{1/2}=6.5$My) and Pd is much more siderophile than Ag, it is relatively straightforward to calculate the conditions under which the $^{107}$Ag/$^{109}$Ag of BSE can be the same as that of CI chondrites. The relatively refractory Pd is generally regarded as having been added to Earth throughout accretion, while moderately volatile Ag could have been added throughout accretion (in a homogeneous accretion scenario) or as part of the Late Veneer. The end-member homogeneous accretion model can be excluded (Schonbachler, et al. 2010) because the apparent age of segregation of Pd from Ag would be, in a 2-stage model, much older than that indicated by the Hf-W system, i.e. 9My instead of 30My. More protracted accretion with a certain fraction of disequilibrium between metal and silicate does not lead to concordance between Hf-W and Pd-Ag “ages” of silicate Earth. The most plausible way of bringing agreement between these two short-lived isotope systems is if the bulk of Earth's Ag had been added after $^{107}$Pd was extinct i.e. more than 30 My after the origin of the solar system. Reducing by dilution the Ag-isotope ratio resulting from Pd decay, however, requires the delivery of 13\% of the mass of the Earth in volatile rich material. This would provide too much Ag relative to the current mantle composition; thus this delivery should have happened while core formation was still on-going (i.e. not in the Late-Veneer phase), which resulted in a large fraction of the added Ag being segregated to the core.

A final argument against the possibility of accretion of volatiles during the sole phase of Late Veneer comes from molybdenum--ruthenium isotopes (Dauphas et al., 2004). In differentiated and bulk primitive meteorites the isotopic anomalies of these two elements correlate with one another, although new data (Chen et al. 2010 and Burkhardt et al. 2011) seem to make the correlation less sharp. Molybdenum is only moderately siderophile, thus most of the amount of Mo presently in the mantle was delivered before the completion of core formation. In contrast, ruthenium is highly siderophile, so that nearly all of the mantle Ru was delivered in the Late Veneer. The fact silicate Earth lies on the Mo--Ru cosmic correlation supports the idea that the Earth accreted quite homogeneously. For instance, if the Earth's mantle had got Mo from enstatite chondrites (dry accretion), and Ru from CM chondrites (wet Late Veneer) then the silicate Earth should be at $\epsilon^{92}$Mo=0 and $\epsilon^{102}$Ru=-1.2, i.e., completely off the correlation, unlike what is observed. 

In summary, although some elementary ratios argue for a volatile-rich Late Veneer (Wang and Becker, 2013) it is unlikely that a dominant fraction of Earth’s volatiles could be delivered by the Late Veneer.

\section{Conclusions}

Late Veneer and Late Accretion are related to different concepts. The first is the delivery of chondritic material that, once on our planet, avoided any metal-silicate segregation. The second is the delivery of material after the last giant impact on Earth. We have shown that Late Veneer and Late Accretion are probably not the same. Some of the HSEs in the Earth’s mantle possibly pre-date the Moon forming event and late-accreted differentiated projectiles probably did not deliver their full HSE budget to the Earth’s mantle. Nevertheless, we estimate that the masses delivered as Late Veneer and Late Accretion are probably within a factor of 2 of each other. 

The small amount of mass accreted by the Earth after the Moon forming event ($<1$\% of the Earth’s mass) implies that the Moon formed during the last accretional giant impact on our planet and that it formed relatively late, approximately 100My after the first solar system solids. It also argues against the possibility that the entire volatile element budget of the Earth was acquired during the Late Veneer. Geochemical and isotopic arguments also rule out this hypothesis. 
The large difference in late accreted masses on the Earth ($\sim 5\times 10^{-3} M_E$) and on the Moon ($5\times 10^{-6} M_E$) suggests that the post-Moon-formation accretion process was dominated by a few large impactors which, by virtue of the large ratio of gravitational cross sections, hit our planet but missed our satellite (Bottke et al., 2010; Raymond et al., 2013).

\acknowledgments 

A.M.. was supported by the European Research Council (ERC) Advanced Grant “ACCRETE” (contract number 290568). The authors wish to thank R.J. Walker and an anonymous reviewer for their constructive reports as well as D. Rubie and S. Jacobson for enriching discussions. 


\section{References}

\begin{itemize}

\item[--] Albarede F (2009) Volatile accretion history of the terrestrial planets and dynamic implications. Nature 461:1227-1233 

\item[--] Albarede, F., Ballhaus, C., Blichert-Toft, J., Lee, C.T., Marty, B., Moynier, F., Yin, Q.Z. (2013) Asteroidal impacts and the origin of terrestrial and lunar volatiles. Icarus 222, 44-52. 
 
\item[--] Allegre, C.J., Manhes, G., Gopel, C. (2008) The major differentiation of the Earth at $\sim$ 4.45 Ga. Earth and Planetary Science Letters 267, 386-398. 
 
\item[--] Anders E (1968) Chemical Processes in Early Solar System, as Inferred from Meteorites. Accounts Chem Res 1(10):289-\& doi:Doi 10.1021/Ar50010a001

\item[--] Bottke, W.F., Levison, H.F., Nesvorny, D., Dones, L. (2007) Can planetesimals left over from terrestrial planet formation produce the lunar Late Heavy Bombardment?. Icarus 190, 203-223. 

\item[--] Bottke, W.F., Walker, R.J., Day, J.M.D., Nesvorny, D., Elkins-Tanton, L. 2010. Stochastic Late Accretion to Earth, the Moon, and Mars. Science 330, 1527. 

\item[--] Bottke, W.F., Vokrouhlicky, D., Minton, D., Nesvorny, D., Morbidelli, A., Brasser, R., Simonson, B., Levison, H.F. (2012) An Archaean heavy bombardment from a destabilized extension of the asteroid belt. Nature 485, 78-81. 

\item[--] Bouvier, A., Wadhwa, M. (2010) The age of the Solar System redefined by the oldest Pb-Pb age of a meteoritic inclusion. Nature Geoscience 3, 637-641.

\item[--] Burkhardt, C., Kleine, T., Oberli, F., Pack, A., Bourdon, B., Wieler, R. 2011. Molybdenum isotope anomalies in meteorites: Constraints on solar nebula evolution and origin of the Earth. Earth and Planetary Science Letters 312, 390-400. 
 
\item[--] Canup, R.M., Asphaug, E. 2001. Origin of the Moon in a giant impact near the end of the Earth's formation. Nature 412, 708-712. 

\item[--] Canup, R.M. 2012. Forming a Moon with an Earth-like Composition via a Giant Impact. Science 338, 1052. 

\item[--] Chen, J.H., Papanastassiou, D.A., Wasserburg, G.J. 2010. Ruthenium endemic isotope effects in chondrites and differentiated meteorites. Geochimica et Cosmochimica Acta 74, 3851-3862. 
 
\item[--] Chou CL (1978) Fractionation of siderophile elements in the earth's upper mantle. In: Lunar and Planetary Science Conference, vol 9. pp 219-230

\item[--] Chou CL, Shaw DM, Crocket JH (1983) Siderophile trace elements in the Earth's oceanic crust and upper mantle. Journa of Geophysical Research 88:A507-A518 

\item[--] Cohen, B.A., Swindle, T.D., Kring, D.A. (2000) Support for the Lunar Cataclysm Hypothesis from Lunar Meteorite Impact Melt Ages. Science 290, 1754-1756.

\item[--] Cuk, M., Stewart, S.T. (2012) Making the Moon from a Fast-Spinning Earth: A Giant Impact Followed by Resonant Despinning. Science 338, 1047. 

\item[--] Dauphas, N., Davis, A.M., Marty, B., Reisberg, L. (2004) The cosmic molybdenum-ruthenium isotope correlation. Earth and Planetary Science Letters 226, 465-475. 
 
\item[--] Dauphas, N., Pourmand, A. (2011) Hf-W-Th evidence for rapid growth of Mars and its status as a planetary embryo. Nature 473, 489-492. 
 
\item[--] Day, J.M.D., Walker, R.J., James, O.B., Puchtel, I.S. (2010) Osmium isotope and highly siderophile element systematics of the lunar crust. Earth and Planetary Science Letters 289, 595-605. 

\item[--] Drake, M.J., Righter, K. 2002. Determining the composition of the Earth. Nature 416, 39-44. 
 
\item[--] Frost DJ, Liebske C, Langenhorst F, McCammon CA, Trønnes RG, Rubie DC (2004) Experimental evidence for the existence of iron-rich metal in the Earth's lower mantle. Nature 428(6981):409-412 

\item[--] Gao X, Thiemens MH (1993) Variations of the Isotopic Composition of Sulfur in Enstatite and Ordinary Chondrites. Geochimica et Cosmochimica Acta 57(13):3171-3176 doi:Doi 10.1016/0016-7037(93)90301-C

\item[--] Georg, R.B., Halliday, A.N., Schauble, E.A., Reynolds, B.C. (2007) Silicon in the Earth's core. Nature 447, 1102-1106. 
 
\item[--] Halliday, A.N., Wood, B.J., Kleine, T. (2005) Runaway Growth of Mars and Implications for Core Formation Relative to Earth. AGU Fall Meeting Abstracts 3. 
 
\item[--] Halliday, A.N. (2008) A young Moon-forming giant impact at 70-110 million years accompanied by late-stage mixing, core formation and degassing of the Earth. Royal Society of London Philosophical Transactions Series A 366, 4163-4181. 
 
\item[--] Herwartz, D. (2013) Differences in the D17O between Earth, Moon and enstatite chondrites. Royal Society/Kavli Institute Meeting on ‘‘The Origin of the Moon— Challenges and Prospects’’ (25–26 September, Chicheley Hall, Buckinghamshire, UK. 

\item[--] Holzheid A, Sylvester P, O'Neill HSC, Rubie DC, Palme H (2000) Evidence for a late chondritic veneer in the Earth's mantle from high-pressure partitioning of palladium and platinum. Nature 406(6794):396-399 doi:Doi 10.1038/35019050

\item[--] Jacobson, S.A., Morbidelli, A., Raymond, S.N., O'Brien, D.P., Walsh, K.J., Rubie, D.C. (2014) Highly siderophile elements in Earth's mantle as a clock for the Moon-forming impact. Nature 508, 84-87. 

\item[--] Jacobson, S.A., Morbidelli, A. (2014) Lunar and Terrestrial Planet Formation in the Grand Tack Scenario. Proceedings of the Royal Society A., in press.
 
\item[--] Kimura K, Lewis RS, Anders E (1974) Distribution of Gold and Rhenium between Nickel-Iron and Silicate Melts - Implications for Abundance of Siderophile Elements on Earth and Moon. Geochimica et Cosmochimica Acta 38(5):683-701 doi:Doi 10.1016/0016-7037(74)90144-6

\item[--] Kleine T, Mezger K, Palme H, Munker C (2004) The W isotope evolution of the bulk silicate Earth: constraints on the timing and mechanisms of core formation and accretion. Earth And Planetary Science Letters 228(1-2):109-123 

\item[--] Kleine, T., Mezger, K., Munker, C., Palme, H., Bischoff, A. (2004b) 182Hf- 182W isotope systematics of chondrites, eucrites, and martian meteorites: 
Chronology of core formation and early mantle differentiation in Vesta and 
Mars. Geochimica et Cosmochimica Acta 68, 2935-2946. 

\item[--] Kleine, T., Kruijer, T.S., Sprung, P. (2014) Lunar 182W and the Age and Origin of the Moon. Lunar and Planetary Science Conference 45, 2895. 
 
\item[--] K\"onig, S., Munker, C., Hohl, S., Paulick, H., Barth, A.R., Lagos, M., Pfander, J., Buchl, A. (2011) The Earth’s tungsten budget during mantle melting and crust formation. Geochimica et Cosmochimica Acta 75, 2119-2136. 

\item[--] Kyser TK, Oneil JR, Carmichael ISE (1986) Possible nonequilibrium oxygen isotope effects in mantle nodules, an alternative to the kyser-oneil-carmichael o-18-o-16 geothermometer - reply. Contributions to Mineralogy and Petrology 93(1):120-123 

\item[--] Labidi J, Cartigny P, Moreira M (2013) Non-chondritic sulphur isotope composition of the terrestrial mantle. Nature 501(7466):208-+ doi:Doi 10.1038/Nature12490

\item[--] Li J, Agee CB (1996) Geochemistry of mantle-core differentiation at high pressure. Nature 381:686-689 

\item[--] Lodders K (2003) Solar system abundances and condensation temperatures of the elements. Astrophys J 591:1220-1247 

\item[--] Lugmair, G.W., Shukolyukov, A. (1998) Early solar system timescales according to 53Mn-53Cr systematics. Geochimica et Cosmochimica Acta 62, 
2863-2886. 
 
\item[--] Maier, W.D., Barnes, S.J., Campbell, I.H., Fiorentini, M.L., Peltonen, P., Barnes, S.J., Smithies, R.H. (2009) Progressive mixing of meteoritic veneer into the early Earth's deep mantle. Nature 460, 620-623. 
 
\item[--] Mann U, Frost DJ, Rubie DC, Becker H, Audetat A (2012) Partitioning of Ru, Rh, Pd, Re, Ir and Pt between liquid metal and silicate at high pressures and high temperatures - Implications for the origin of highly siderophile element concentrations in the Earth's mantle. Geochimica et Cosmochimica Acta 84:593-613 doi:Doi 10.1016/J.Gca.2012.01.026

\item[--] Marchi, S., Bottke, W.F., Kring, D.A., Morbidelli, A. (2012) The onset of the lunar cataclysm as recorded in its ancient crater populations. Earth and Planetary Science Letters 325, 27-38. 

\item[--] Marchi, S., Bottke, W.F., Elkins-Tanton, L.T., Bierhaus, M.,  Wuennemann, K.,  Morbidelli, A.,  Kring, D.A. (2014) Widespread mixing and burial of Earth's Hadean crust by asteroid impacts. Submitted.

\item[--] McDonough WF, Sun S-s (1995) The composition of the Earth. Chemical Geology 120(3-4):223-253 

\item[--] Morbidelli, A., Tsiganis, K., Crida, A., Levison, H.F., Gomes, R. (2007) Dynamics of the Giant Planets of the Solar System in the Gaseous Protoplanetary Disk and Their Relationship to the Current Orbital Architecture. The Astronomical Journal 134, 1790-1798. 

\item[--] Morbidelli, A., Marchi, S., Bottke, W.F., Kring, D.A. (2012) A sawtooth-like timeline for the first billion years of lunar bombardment. Earth and Planetary Science Letters 355, 144-151.

\item[--] Morbidelli, A. (2013) Dynamical Evolution of Planetary Systems. Planets, Stars and Stellar Systems. Volume 3: Solar and Stellar Planetary Systems 63.

\item[--] Neumann, G.A., and 10 colleagues (2013) The Inventory of Lunar Impact Basins from LOLA and GRAIL. Lunar and Planetary Science Conference 44, 2379.

\item[--] O'Neill HS (1991) The Origin of the Moon and the Early History of the Earth - a Chemical-Model .2. The Earth. Geochimica et Cosmochimica Acta 55(4):1159-1172 doi:Doi 10.1016/0016-7037(91)90169-6

\item[--] Palme H, O'Neill HSC (2003) Cosmochemical Estimates of Mantle Composition. In: Carlson RW (ed) The Mantle and Core, vol 2. Elsevier, Amsterdam, pp 1-38

\item[--] Raymond, S.N., Schlichting, H.E., Hersant, F., Selsis, F. (2013) Dynamical and collisional constraints on a stochastic late veneer on the terrestrial 
planets. Icarus 226, 671-681. 

\item[--] Reufer, A., Meier, M.M.M., Benz, W., Wieler, R. 2012. A hit-and-run giant impact scenario. Icarus 221, 296-299. 

\item[--] Righter K, Humayun M, Danielson L (2008) Partitioning of palladium at high pressures and temperatures during core formation. Nature Geoscience 1(5):321-323 doi:Doi 10.1038/Ngeo180

\item[--] Rubie, D.C., Frost, D.J., Mann, U., Asahara, Y., Nimmo, F., Tsuno, K., Kegler, P., Holzheid, A., Palme, H. (2011) Heterogeneous accretion, composition and core-mantle differentiation of the Earth. Earth and Planetary Science Letters 301, 31-42. 

\item[--] Ryder, G. (2002) Mass flux in the ancient Earth-Moon system and benign implications for the origin of life on Earth. Journal of Geophysical Research (Planets) 107, 5022. 

\item[--] Schonbachler M, Carlson RW, Horan MF, Mock TD, Hauri EH (2010) Heterogeneous accretion and the moderately volatile element budget of Earth. Science 328:884-887 

\item[--] Stoffler, D., Ryder, G. (2001) Stratigraphy and Isotope Ages of Lunar Geologic Units: Chronological Standard for the Inner Solar System. Space Science Reviews 96, 9-54. 

\item[--] Strom, R.G., Malhotra, R., Ito, T., Yoshida, F., Kring, D.A. (2005) The Origin of Planetary Impactors in the Inner Solar System. Science 309, 1847-1850. 

\item[--] Taylor, D.J., McKeegan, K.D., Harrison, T.M., Young, E.D. (2009) Early differentiation of the lunar magma ocean. New Lu-Hf isotope results from Apollo 17. Geochimica et Cosmochimica Acta Supplement 73, 1317. 
 
\item[--] Tera, F., Papanastassiou, D.A., Wasserburg, G.J. (1974) Isotopic evidence for a terminal lunar cataclysm. Earth and Planetary Science Letters 22, 1-21. 

\item[--] Thibault Y, Walter MJ (1995) The influence of pressure and temperature on the metal-silicate partition coefficients of nickel and cobalt in a model C1 chondrite and implications for metal segregation in a deep magma ocean. Geochimica Et Cosmochimica Acta 59(5):991-1002 

\item[--] Touboul M, Kleine T, Bourdon B, Palme H, Wieler R (2007) Late formation and prolonged differentiation of the Moon inferred from W isotopes in lunar metals. Nature 450:1206-1209 

\item[--] Touboul, M,, Puchtel, I.S.,  Walker, R.J. (2012)  182W Evidence for Long-Term
Preservation of Early Mantle Differentiation Products. Science, 335, 1065-1069

\item[--] Turekian KK, Clark SP (1969) Inhomogeneous Accumulation of Earth from Primitive Solar Nebula. Earth and Planetary Science Letters 6(5):346-\& doi:Doi 10.1016/0012-821x(69)90183-6

\item[--] Wade J, Wood BJ (2005) Core formation and the oxidation state of the Earth. Earth And Planetary Science Letters 236:78-95 

\item[--] Walker R.J., Horan MF, Morgan JW, Becker H, Grossman JN, Rubin AE (2002) Comparative 187Re-187Os systematics of chondrites: Implications regarding early solar system processes. . Geochimica et Cosmochimica Acta 66:4187-4201 

\item[--] Walker, R.J. (2009) Highly siderophile elements in the Earth, Moon and Mars: Update and implications for planetary accretion and differentiation. Chemie der Erde Geochemistry 69, 101-125. 
 
\item[--] Wang ZC, Becker H (2013) Ratios of S, Se and Te in the silicate Earth require a volatile-rich late veneer. Nature 499(7458):328-+ doi:Doi 10.1038/Nature12285

\item[--] Wiechert U, Halliday AN, Lee D-C, Snyder GA, Taylor LA, Rumble DA (2001) Oxygen isotopes and the Moon-forming giant impact. Science 294:345-348 

\item[--] Wilhelms, D.E., (1987). The geologic history of the Moon. (U.S. Geol. Surv. Prof. Pap., 1348).

\item[--] Willbold M, Elliott T, Moorbath S (2011) The tungsten isotopic composition of the Earth's mantle before the terminal bombardment. Nature 477(7363):195-U191 doi:Doi 10.1038/Nature10399

\item[--] Wood BJ, Halliday A, Rehkamper M (2010) Volatile accretion history of the Earth. Nature 467:E6-E7 

\item[--] Wood BJ, Wade J, Kilburn MR (2008) Core formation and the oxidation state of the Earth: Additional constraints from Nb, V and Cr partitioning. Geochimica et Cosmochimica Acta 72:1415-1426 

\item[--] Yin Q.Z., Jacobsen SB, Yamashita K, Blichert-Toft J, Telouk P, Albarede F (2002) A short timescale for terrestrial planet formation from Hf-W chronometry of meteorites. Nature 418(6901):949-952 

\item[--] Yin, Q.Z., Antognini, J. (2008) Isotopic and elemental constraints on the first 100 Myr of Earth history. Geochimica et Cosmochimica Acta Supplement 72, 1061. 

\item[--] Zhang, J., Dauphas, N., Davis, A.M., Leya, I., Fedkin, A. (2012) The proto-Earth as a significant source of lunar material. Nature Geoscience 5, 251-255.

\end{itemize}

\end{document}